%% file: Template.tex
\documentclass{Interspeech}

\interspeechcameraready

\usepackage{amsmath,graphicx}
\usepackage{amsfonts}
\usepackage{booktabs}
\usepackage[subrefformat=parens]{subcaption}
\usepackage{siunitx}
\usepackage{mathtools}
\usepackage{comment}
\usepackage{multirow}
\usepackage{makecell}
\usepackage{arydshln}   
\usepackage{cite}

\usepackage{tikz}
\usetikzlibrary{calc}
\usetikzlibrary{tikzmark}
\usetikzlibrary{arrows, positioning,intersections,angles}
\usetikzlibrary{fit}
\usetikzlibrary{decorations.pathreplacing}
\usetikzlibrary{patterns,patterns.meta}

\def\equationautorefname~#1\null{(#1\null)}

\renewcommand{\sectionautorefname}{Section}
\renewcommand{\subsectionautorefname}{\sectionautorefname}

\let\orgautoref\autoref
\providecommand{\Autoref}[1]
{%
\def\figureautorefname{Figure}%
\def\subfigureautorefname{Figure}%
\orgautoref{#1}%
}
\renewcommand{\autoref}[1]
{%
\def\figureautorefname{Fig.}%
\def\subfigureautorefname{\figureautorefname}%
\def\sectionautorefname{Sec.}%
\def\subsectionautorefname{\sectionautorefname}%
\def\subsectionautorefname{\sectionautorefname}%
\orgautoref{#1}%
}

\newcommand{\vect}[1]{\boldsymbol{#1}}
\newcommand{\abs}[1]{\left\lvert#1\right\rvert}

\newcommand{\trans}[1]{#1^\mathsf{T}}

\def\appendixautorefname~#1\null{~#1 \null}

\newcommand{\shortdots}{.\mkern2mu.\mkern2mu.}

\newcommand{\svss}{\textit{one-vs-one}}
\newcommand{\svsm}{\textit{one-vs-many}}
\newcommand{\system}[2]{\texttt{#1}-\texttt{#2}}

\makeatletter

\newcommand{\tblcaption}[1]{\captionsetup{type=table}\caption{#1}}
\makeatother

\makeatletter
   {\list{}{\leftmargin=10pt 
            \labelwidth\z@ \itemindent-\leftmargin
            }}%
   {\endlist}
\makeatother

\newcommand{\customdashline}[1]{%
  \noalign{\vskip\aboverulesep} 
  \cdashline{#1}
  \noalign{\vskip\belowrulesep} 
}

\title{Mitigating Non-Target Speaker Bias in Guided Speaker Embedding}

\author[]{Shota}{Horiguchi}
\author[]{Takanori}{Ashihara}
\author[]{Marc}{Delcroix}
\author[]{Atsushi}{Ando}
\author[]{Naohiro}{Tawara}

\affiliation[nocounter]{}{NTT, Corporation.}{Japan}

\email{horiguchi@ieee.org}

\begin{document}
\abovedisplayskip=1pt
\belowdisplayskip=1pt

\setlength\textfloatsep{5pt}
\setlength\dbltextfloatsep{5pt}
\setlength\abovecaptionskip{2pt}
\setlength\belowcaptionskip{2pt}

\maketitle
\begin{abstract}
Obtaining high-quality speaker embeddings in multi-speaker conditions is crucial for many applications.
A recently proposed guided speaker embedding framework, which utilizes speech activities of target and non-target speakers as clues, drastically improved embeddings under severe overlap with small degradation in low-overlap cases.
However, since extreme overlaps are rare in natural conversations, this degradation cannot be overlooked.
This paper first reveals that the degradation is caused by the global-statistics-based modules, widely used in speaker embedding extractors, being overly sensitive to intervals containing only non-target speakers.
As a countermeasure, we propose an extension of such modules that exploit the target speaker activity clues, to compute statistics from intervals where the target is active.
The proposed method improves speaker verification performance in both low and high overlap ratios, and diarization performance on multiple datasets.
\end{abstract}

\section{Introduction}
In long-form multi-speaker audio processing, a commonly used approach involves processing short segments to obtain results with intra-segment speaker labels and speech activity intervals, followed by aligning speakers across segments.
For example, some speaker diarization methods first estimate speaker-wise speech activities within a sliding window and then estimate which pair of speakers across windows are the same identity~\cite{kinoshita2021integrating,bredin2023pyannote,plaquet2023powerset}.
The same can also be done for speech recognition methods that provide overlap-aware multi-speaker recognition results along with intra-segment speaker labels and timestamps~\cite{lu2021streaming,lu2022endpoint,sklyar2022multi,kanda2022streaming,cui2024improving,moriya2025alignment}.
This two-stage approach has the advantage of eliminating the need to develop models that can handle an infinite number of speakers and input lengths.

Speaker embeddings are crucial for aligning speakers across segments, as they determine whether speakers across segments are the same or different based on their similarities.
Here, it is important that the speaker embeddings do not get contaminated by interference speakers.
One commonly used approach is to compute the embedding using only intervals where each speaker is speaking alone~\cite{kanda2019simultaneous,medennikov2020targetspeaker,kinoshita2021integrating,bredin2023pyannote,plaquet2023powerset}.
However, it may result in low-quality embeddings when the single-speaker intervals are too short or even not available.\footnote{In such cases, speaker embeddings are often extracted from overlapping segments out of necessity~\cite{bredin2023pyannote}.}
Another approach is to use speech separation or extraction to remove interference speakers beforehand~\cite{delcroix2021speaker}, but they usually cause artifacts and distortion that may affect the quality of the embeddings.
To address these limitations, guided speaker embedding has been recently proposed to extract speaker embeddings directly from multi-speaker recordings by leveraging speech activities of target and non-target speakers~\cite{horiguchi2025guided}.
This is a promising direction as it enables speaker embedding extraction from overlapping segments without relying on separation or enhancement.
However, the performance in situations with less overlap is slightly compromised.
Since the typical overlap ratio in real-world conversations is not very high, there are cases where the significant improvement in highly overlapping situations cannot outweigh the degradation in less overlapping situations.

\begin{figure}[t]
\centering
\input{figs/diagram.tex}
\caption{Block diagram of standard single-speaker embedding extraction, guided speaker embedding extraction, and the proposed bias-mitigated guided speaker embedding extraction.}\label{fig:diagram}
\end{figure}
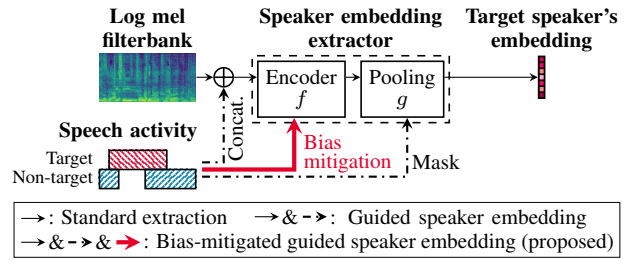

This paper first identifies the cause of the aforementioned performance degradation in guided speaker embedding.
While non-target speaker intervals help identify the information to be removed in overlapping speech, they also introduce unnecessary bias, which becomes apparent when the overlap is minimal.
We reveal that this bias arises in widely-used modules that use statistics computed over the entire input sequence, caused by intervals where only non-target speakers speak, contaminating these statistics.
In the proposed method, we mitigate this bias by adapting the modules to use statistics calculated solely from the target speaker's intervals, instead of relying on global statistics as in \autoref{fig:diagram}.
To demonstrate the broad effectiveness of the proposed method, experimental evaluation is performed using two different architectures (ECAPA-TDNN~\cite{desplanques2020ecapatdnn} and CAM++~\cite{wang2023cam}) on two different tasks (speaker verification and diarization).

\section{Review of guided speaker embedding}
\subsection{Method overview}\label{sec:guided_overview}
The guided speaker embedding framework enables the extraction of the target speaker's embedding, specified by speech activities of target and non-target speakers~\cite{horiguchi2025guided}.
This section provides the formulation of the framework.

We first introduce a common speaker embedding extractor.
Let $\left[\vect{x}_t\right]_{t=1}^T$ denote a sequence of acoustic features $\vect{x}_t\in\mathbb{R}^F$, where $t$ is the time index.
The processes inside the extractor to compute the speaker embedding $\vect{v}$ are as follows:
\begin{align}
    [\vect{h}_t]_{t=1}^T&=f\big(\left[\vect{x}_t\right]_{t=1}^T\big)\in\mathbb{R}^{D\times T},\label{eq:encoder}\\
    \vect{v}&=g\big([\vect{h}_t]_{t=1}^T\big)\in\mathbb{R}^{E}.\label{eq:pooling}
\end{align}
$f$ is the encoder that transforms input features into frame-wise embeddings $[\vect{h}_t]_{t=1}^T$, and $g$ is the pooling mechanism to aggregate the frame-wise embeddings into a single embedding, followed by dimensionality reduction with a linear layer.

In guided speaker embedding, the following two modifications are made to the standard extractor as shown in \autoref{fig:diagram}.
The first modification is that speech activities of target and non-target speakers are used for conditioning the input to the encoder, i.e., the following replaces the process in \autoref{eq:encoder}:
\begin{equation}
    [\vect{h}_t]_{t=1}^T=f\left(\left[\vect{x}_t\oplus\trans{\left[q^{\text{target}}_t,q^{\text{non-target}}_t\right]}\right]_{t=1}^T\right)\in\mathbb{R}^{D\times T},\label{eq:encoder_guided}
\end{equation}
where $\oplus$ denotes vector concatenation, $q^{\text{target}}_t$ denotes the target speaker's speech activity that takes $1$ if the target speaker is active at $t$ and $0$ otherwise, and $q^{\text{non-target}}_t$ takes $1$ if at least one non-target speaker is active at $t$ and $0$ otherwise.
This enables enhancing the target speaker's information while suppressing that of non-target speakers in the encoder.
The second modification is to use only the embeddings of frames in which the target speaker is speaking for pooling:
\begin{equation}
    \vect{v}=g\left([\vect{h}_t\mid q_t^{\text{target}}=1]\right)\in\mathbb{R}^{E}.\label{eq:pooling_guided}
\end{equation}
Removing inactive intervals of the target speaker prevents contamination from non-target speakers in the output embedding.

\subsection{Remaining problem}\label{sec:remaining_problem}
The important problem of guided speaker embedding is, as reported in the previous study, the degradation in less-overlapping conditions.
The degree of degradation is indeed small (0.2 points in equal error rate, EER, in the single-speaker case~\cite{horiguchi2025guided}), but it causes a non-negligible performance drop in real applications, e.g., speaker diarization, since the overlap ratio of real conversations is typically not very high.

\begin{figure}[t]
    \includegraphics[width=\linewidth]{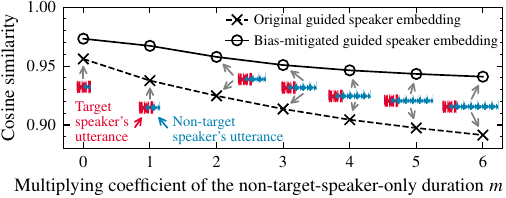}
    \caption{Cosine similarity between embeddings extracted using guided speaker embedding in a two-speaker case: one from a target speaker's utterance and the other from a target/non-target mixture, with varying non-target-speaker-only durations.}
    \label{fig:cosine_similarity}
\end{figure}

To reveal the reason for this, \autoref{fig:cosine_similarity} shows an example of a two-speaker case: a target and a non-target speaker.
We computed the cosine similarity between speaker embeddings extracted from a target speaker's single-speaker utterance and its mixture with a non-target speaker's utterance using the ECAPA-TDNN-based guided speaker embedding system~\cite{horiguchi2025guided}.
We augmented the non-target-speaker-only duration by repeating the corresponding segment as in \autoref{fig:cosine_similarity}, where $m=1$ is the original mixture, $m=0$ removes non-target-speaker-only intervals, and $m>1$ extends them $m$-fold.
As in \autoref{sec:guided_overview}, guided speaker embedding is intentionally designed to convey no information about non-target speakers.
However, lengthening the non-target speaker's interval obviously affected the output embedding as the cosine similarity decreased.
The proposed method, explained in the next section, successfully mitigates the decrease in cosine similarity with increasing $m$, as shown in \autoref{fig:cosine_similarity}.

\section{Proposed method}\label{sec:proposed}
In this paper, we extend the conditioning introduced in guided speaker embedding to all modules computing global statistics.
This is based on the hypothesis that the excessive influence of non-target speakers, which occurs not only during inference as in \autoref{sec:remaining_problem} but also during training, stems from modules using global statistics, and we propose using statistics calculated from the target speaker's intervals instead.
In particular, we focus on the ECAPA-TDNN~\cite{desplanques2020ecapatdnn} and CAM++~\cite{wang2023cam} architectures and describe how the proposed method is applied to squeeze-and-excitation (SE) block, context-aware masking, and batch normalization used therein.
Note that the proposed method can be applied to other extractors that use these modules~\cite{koluguri2022titanet,zhao2023pcf,liu2023ecapa++}, those that employ other modules based on global statistics~\cite{thienpondt2021integrating,liu2022mfa,heo2024next}.

\subsection{Guided squeeze-and-excitation block}
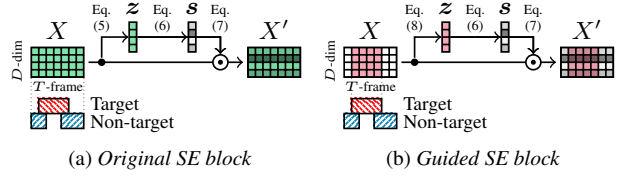
\begin{figure}
\centering
\subfloat[Original SE block]{\scalebox{1.04}{\input{figs/basic_se.tex}}\label{fig:se_block}}\hfill
\subfloat[Guided SE block]{\scalebox{1.04}{\input{figs/target_se.tex}}\label{fig:guided_se_block}}
\caption{Original and proposed guided SE blocks. The colored cells in $X$ are used for computing \(\vect{z}\).}
\end{figure}

SE block is a module to recalibrate channel-wise importance~\cite{hu2018squeeze}.
After being first proposed for computer vision models, it was introduced into speaker verification models and is now used in various architectures~\cite{zhou2021resnext,desplanques2020ecapatdnn,wang2022efficienttdnn,zhao2023pcf}.
In this paper, we focus on a one-dimensional variant of SE blocks, which is adopted in widely-used ECAPA-TDNN~\cite{desplanques2020ecapatdnn}.

In ECAPA-TDNN, the SE block is inserted just after each TDNN-Res2Net-TDNN chain.
Let $X=\left[\vect{x}_t\right]_{t=1}^{T}\in\mathbb{R}^{D\times T}$ be an input $T$-length sequence of $D$-dimensional features.
The squeeze operation is a global average pooling for each channel:
\begin{equation}
    \vect{z}=\frac{1}{T}\sum_{t=1}^T \vect{x}_t.\label{eq:squeeze}
\end{equation}
The excitation operation is channel-wise weighting using two-stacked linear layers:
\begin{align}
    \vect{s}&=\sigma\left(W_4\delta\left(W_3\vect{z}+\vect{b}_3\right)+\vect{b}_4\right)\in\left(0,1\right)^D,\label{eq:weight}\\[-2pt]
    X'&=X\odot[\underbrace{\vect{s},\dots,\vect{s}}_{T}]\in\mathbb{R}^{D\times T},\label{eq:excitation}
\end{align}
where $\{W_3\in\mathbb{R}^{\frac{D}{r}\times D}, \vect{b}_3\in\mathbb{R}^{\frac{D}{r}}\}$ and $\{W_4\in\mathbb{R}^{D\times\frac{D}{r}}, \vect{b}_4\in\mathbb{R}^{D}\}$ are the parameters of linear layers, $r\in\mathbb{R}_{>0}$ is the dimensionality reduction ratio, $\odot$ denotes element-wise product, and $\sigma\left(\cdot\right)$ and $\delta\left(\cdot\right)$ are sigmoid and rectified linear unit activations.
These processes inside the SE block are depicted in \autoref{fig:se_block}.

The original SE block uses the entire sequence to compute channel-wise weights in \autoref{eq:squeeze}--\autoref{eq:weight}.
However, as shown in \autoref{fig:cosine_similarity}, the performance of the guided speaker embedding decreases as the duration of non-target speaker intervals increases.
This is likely because non-target speaker intervals dominate the calculation of weights in an SE block, making it difficult to emphasize channels that are effective in extracting the target speaker's information.
To mitigate this, we propose calculating channel-wise weights using only frames in which the target speaker is active $\mathcal{Q^\text{target}}\coloneqq\{t\mid q_t^\text{target}=1\}$ as shown in \autoref{fig:guided_se_block}.
Specifically, the squeeze operation in \autoref{eq:squeeze} is replaced by
\begin{align}
    \vect{z}=\frac{1}{\abs{\mathcal{Q}^\text{target}}}\sum_{t\in\mathcal{Q}^\text{target}} \vect{x}_t.\label{eq:guided_z}
\end{align}
This allows the SE block to place more weight on channels that are important for extracting the target speaker's information.

\subsection{Guided context-aware masking}
Context-aware masking (CAM) can be seen as an extension of SE~\cite{yu2021cam}.
While the SE block directly changes the input $X$ via channel-wise weighting, CAM performs input transformation $g:\mathbb{R}^{D\times T }\rightarrow\mathbb{R}^{D\times T}$ using TDNNs and channel-wise weight computation from the input by \autoref{eq:weight} in parallel.
The output is calculated as their channel-wise product by replacing \autoref{eq:excitation} with
\begin{equation}
    X'=g(X)\odot [\underbrace{\vect{s},\dots,\vect{s}}_{T}]\in\mathbb{R}^{D\times T}.\label{eq:cam}
\end{equation}

In this paper, we use the enhanced version of CAM, namely, CAM++,~\cite{wang2023cam}.
The original CAM calculates the weights using global statistics $z$, but local statistics are also used in CAM++.
Specifically, a $T$-length input is split into $K$ segments, where the indices of the $k$-th segment is $\left\{t_{k-1}+1,t_{k-1}+2,\dots,t_k\right\}$ with $t_0=0$ and $t_K=T$.
Instead of \autoref{eq:weight} and \autoref{eq:cam}, the weighting is based on not only $z$ in \autoref{eq:squeeze} but also on the local statistics $\vect{z}_k$:\footnote{We did not use the proposed guiding method for $\vect{z}_k$ since it is uncomputable in local segments where the target speaker is not speaking.}
\begin{align}
    \vect{z}_k&=\frac{1}{t_k-t_{k-1}}\sum_{t=t_{k-1}+1}^{t_k} \vect{x}_t,\label{eq:local_squeeze}\\[-3pt]
    \vect{s}_k&=\sigma\left(W_4 \delta\left(W_3\left(\vect{z}+\vect{z}_k\right)+\vect{b}_3\right)+\vect{b}_4\right)\in\left(0,1\right)^D,\label{eq:weights_campplus}\\
    X'&=g(X)\odot [\underbrace{\vect{s}_1,\shortdots,\vect{s}_1}_{t_1},\underbrace{\vect{s}_2,\shortdots,\vect{s}_2}_{t_2-t_1},\shortdots,\underbrace{\vect{s}_K,\dots,\vect{s}_K}_{T-t_{K-1}}].
\end{align}

Since CAM++ also uses global statistics $\vect{z}$ in \autoref{eq:weights_campplus}, it faces the same issue as the SE block, where the importance of channels is excessively biased toward non-target speakers.
The proposed method, namely guided CAM++, replaces $z$ with the one calculated only from the target speaker's intervals using \autoref{eq:guided_z}.

\subsection{Guided batch normalization}
Batch normalization~\cite{ioffe2015batch} smooths the loss function in deep neural networks, preventing vanishing or exploding gradients and enabling fast convergence with a large learning rate~\cite{bjorck2018understanding,santurkar2018does}.
Assume a mini-batch of input $\vect{X}=[X_b]_{b=1}^B$, where $B$ is the batch size and
$X_b=\left[x_{b,d,t}\right]_{d=1,t=1}^{D,T}$
denotes the $b$-th sample---$T$-length sequence of $D$-dimensional vectors.
Batch normalization transforms each sample $X_b$ into the same size of matrix $Y_b=\left[y_{b,d,t}\right]_{d=1,t=1}^{D,T}$ by
\begin{equation}
    y_{b,d,t}=\gamma_d\frac{x_{b,d,t}-\mathbb{E}_d(\vect{X})}{\sqrt{\mathbb{V}_d(\vect{X})+\varepsilon}}+\beta_d,
\end{equation}
where $\mathbb{E}_d\left(\cdot\right)$ and $\mathbb{V}_d\left(\cdot\right)$ denote mean and variance of all the elements of the $d$-th dimension of input tensor, i.e.,
\begin{align}
    \mathbb{E}_d(\vect{X})&=\frac{1}{BT}\sum_{b=1}^B\sum_{t=1}^T x_{b,d,t},\label{eq:bn_mean}\\[-3pt]
    \mathbb{V}_d(\vect{X})&=\frac{1}{BT}\sum_{b=1}^B\sum_{t=1}^T \left(x_{b,d,t}-\mathbb{E}_d(\vect{X})\right)^2,\label{eq:bn_variance}
\end{align}
$\vect{\gamma}=\trans{\left[\gamma_1,\dots,\gamma_D\right]}\in\mathbb{R}^D$ and $\vect{\beta}=\trans{\left[\beta_1,\dots,\beta_D\right]}\in\mathbb{R}^D$ are learnable parameters, and $\varepsilon\ll1$ is for numerical stability.

In guided speaker embedding, target-speaker intervals mainly contribute to the output speaker embedding.
If batch normalization is applied across all frames, mean and variance are also affected by those in which the target speaker is not active; this could unnecessarily bias the normalized results toward non-target speakers.
We propose using the statistics calculated only from target-speaker intervals by replacing \autoref{eq:bn_mean}--\autoref{eq:bn_variance} with
\begin{align}
    \mathbb{E}_d(\vect{X})&=\frac{1}{B\abs{\mathcal{Q}^\text{target}}}\sum_{b=1}^B\sum_{t\in\mathcal{Q}^\text{target}} x_{b,d,t},\\[-3pt]
    \mathbb{V}_d(\vect{X})&=\frac{1}{B\abs{\mathcal{Q}^\text{target}}}\sum_{b=1}^B\sum_{t\in\mathcal{Q}^\text{target}} \left(x_{b,d,t}-\mathbb{E}_d(\vect{X})\right)^2.
\end{align}
Note that guided batch normalization is effective only during training and, during inference, uses the mean and variance computed during training, like standard batch normalization.

\section{Experiments}
\subsection{Experimental setup}
The following two encoders were used for the speaker embedding extractors: ECAPA-TDNN with 1024 channels~\cite{desplanques2020ecapatdnn} and CAM++~\cite{wang2023cam}.
We used channel- and context-dependent attentive statistics pooling~\cite{desplanques2020ecapatdnn} to aggregate frame-wise embeddings from the encoders, followed by a linear layer to compute a 192-dimensional output.
Each model takes 80-dimensional log mel filterbanks extracted with a \SI{25}{\ms} window and a \SI{10}{\ms} shift as input.
In the case of guided speaker embedding, target and non-target speakers' speech activities are also input.

The models were trained using the VoxCeleb1\&2 development set~\cite{nagrani2020voxceleb}.
The baseline single-speaker models were trained using utterances cropped to three seconds with a batch size of 256.
The guided speaker embedding models were trained using 256 partially overlapped three-speaker mixtures, resulting in a batch size of 768.
These mixtures are created on the fly by adding 2–4 second utterance clips, each shifted by at least 0.5 seconds.
Each model was trained using additive angular margin softmax loss~\cite{deng2022arcface} with a margin of 0.2 and scale of 30.
We use the Adam optimizer~\cite{kingma2015adam} for 80 epochs of training with cyclical learning rate scheduling.
Each cycle lasts 20 epochs, warming up in the first 1\,000 iterations and decaying in the rest using cosine annealing.
On-the-fly data augmentation using noise~\cite{snyder2015musan} and room impulse responses~\cite{ko2017study} were applied during training.

When the proposed method is applied to ECAPA-TDNN, the batch normalization and SE block are replaced by the guided variants described in \autoref{sec:proposed}.
In the case of CAM++, the batch normalization in dense blocks and context-aware masking are replaced by the corresponding variants, 

\subsection{Speaker verification results}

\begin{figure*}[t]
    \begin{minipage}{0.65\linewidth}
    \tblcaption{Speaker verification results in EERs (\%). Single-speaker models (\system{B1}{1} \& \system{B2}{1}) used only single-speaker intervals for extraction. \dag/\ddag~denotes that \system{P1}{1} (\system{P2}{1}) is significantly different from \system{B1}{1}/\texttt{2} (\system{B2}{1}/\texttt{2}) with bootstrap method ($p$\textless .05, $n$=1000).}\label{tbl:result_veri}
    \centering
    \sisetup{detect-weight,mode=text}
    \setlength{\tabcolsep}{2pt}
    \renewrobustcmd{\bfseries}{\fontseries{b}\selectfont}
    \renewrobustcmd{\boldmath}{}
    \newrobustcmd{\B}{\bfseries}
    \resizebox{\linewidth}{!}{%
    \begin{tabular}{@{}lllS[table-format=1.2]*{5}{S[table-format=1.2]}@{}}
        \toprule
        &&&\svss&\multicolumn{5}{c@{}}{\svsm}\\\cmidrule(l{\tabcolsep}r{\tabcolsep}){4-4}\cmidrule(l{\tabcolsep}){5-9}
        &ID&Method&{\SI{0}{\percent}}&{(0,25)\,\si{\percent}}&{[25,50)\,\si{\percent}}&{[50,75)\,\si{\percent}}&{[75,100)\,\si{\percent}}&{\SI{100}{\percent}}\\\midrule
        \multicolumn{5}{@{}l}{\textbf{Encoder: ECAPA-TDNN}}\\
        \multirow{2}{*}{\begin{tabular}{@{}c@{}}\rotatebox[origin=c]{90}{\scalebox{0.9}{\scriptsize 
Baseline}}\end{tabular}}&\system{B1}{1}&Single-speaker model~\cite{desplanques2020ecapatdnn} & \B 0.88 & 1.29 & 1.75 & 2.59 & 14.37 & 27.64\\
        &\system{B1}{2}& Guided speaker embedding~\cite{horiguchi2025guided}& 1.10 & 1.23 & 1.33 & 1.58 & 2.85 & \B 7.37\\\customdashline{1-9}
        \multirow{3}{*}{\begin{tabular}{@{}c@{}}\rotatebox[origin=c]{90}{\scalebox{0.9}{\scriptsize Proposed}}\end{tabular}}&\system{P1}{1}& Bias-mitigated guided speaker embedding & 0.93\rlap{$^\ddag$} & \B 1.02\rlap{$^{\dag\ddag}$} & \B 1.18\rlap{$^{\dag\ddag}$} & \B 1.38\rlap{$^{\dag\ddag}$} & \B 2.37\rlap{$^{\dag\ddag}$} & 8.09\rlap{$^{\dag\ddag}$}\\
        &\system{P1}{2}& $\hookrightarrow$ w/o guided batch normalization & 0.90 & 1.04 & 1.24 & 1.42 & 2.56 & 8.94\\
        &\system{P1}{3}& $\hookrightarrow$ w/o guided SE block & 1.32 & 1.63 & 1.58 & 1.86 & 3.19 & 7.82\\\midrule
        \multicolumn{5}{@{}l}{\textbf{Encoder: CAM++}}\\
        \multirow{2}{*}{\begin{tabular}{@{}c@{}}\rotatebox[origin=c]{90}{\scalebox{0.9}{\scriptsize Baseline}}\end{tabular}}&\system{B2}{1}&Single-speaker model~\cite{wang2023cam}& \B 0.76 & 1.13 & 1.56 & 2.50 & 13.84 & 28.70\\
        &\system{B2}{2}& Guided speaker embedding~\cite{horiguchi2025guided}& 0.93 & 1.03 & 1.15 & 1.44 & 2.66 & 7.68\\\customdashline{1-9}
        \multirow{3}{*}{\begin{tabular}{@{}c@{}}\rotatebox[origin=c]{90}{\scalebox{0.9}{\scriptsize Proposed}}\end{tabular}}&\system{P2}{1}& Bias-mitigated guided speaker embedding & 0.88\rlap{$^\dag$} & \B 0.94\rlap{$^\dag$} & \B 1.04\rlap{$^{\dag\ddag}$} & \B 1.21\rlap{$^{\dag\ddag}$} & 2.13\rlap{$^{\dag\ddag}$} & 6.69\rlap{$^{\dag\ddag}$}\\
        &\system{P2}{2}& $\hookrightarrow$ w/o guided batch normalization & 0.88 & 0.99 & 1.07 & 1.28 & \B 2.10 & \B 6.40\\
        &\system{P2}{3}& $\hookrightarrow$ w/o guided CAM++ & 0.90 & 1.07 & 1.19 & 1.38 & 2.59 & 7.23\\
        \bottomrule
    \end{tabular}%
    }
    \end{minipage}\hfill
    \begin{minipage}{0.33\linewidth}
        \includegraphics[width=\linewidth]{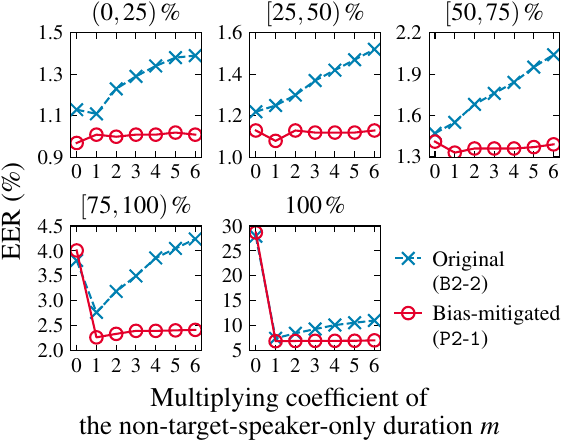}
        \caption{EERs with varied duration where the target speaker is not speaking. The multiplication coefficient $m=1$ corresponds to the original one-vs-many evaluation set.}
        \label{fig:bias_mitigation}
        \vspace{-1em}
    \end{minipage}
\end{figure*}

Following the original guided speaker embedding study~\cite{horiguchi2025guided}, we used VoxCeleb1-O for the standard speaker verification between single-speaker utterances (\svss{}) and its extended versions for verifying whether the target speaker in a multi-speaker mixture is the same identity in a single-speaker utterance  (\svsm{}).
For the latter purpose, each trial in VoxCeleb1-O was extended by mixing three interference speakers' utterances, resulting in five evaluation sets with different overlap ratios of the target speaker's utterance: (0,25)\,\si{\percent}, [25,50)\,\si{\percent}, [50,75)\,\si{\percent}, [75,100)\,\si{\percent}, and \SI{100}{\percent}, i.e., fully overlapped.
We report EERs assuming that oracle speaker-wise speech activities are available.

\Autoref{tbl:result_veri} shows the results of speaker verification.
First, we focus on the results of ECAPA-TDNN.
In line with the previous study~\cite{horiguchi2025guided}, the original guided speaker embedding (\system{B1}{2}) improved EERs on the \svsm{} condition compared to the single-speaker model (\system{B1}{1}), while degraded on the \svss{} condition.
The proposed bias-mitigated guided speaker embedding (\system{P1}{1}) not only improved the EER in \svss{} from \SI{1.10}{\percent} to \SI{0.93}{\percent}, but also enhanced the EER in all \svsm{} conditions except for the fully overlapped one.
This is the expected behavior of the proposed method, which improves speaker discrimination in low-overlapping conditions. 
We also performed an ablation study by replacing guided batch normalization and SE blocks with their original versions in \system{P1}{2} and \system{P1}{3}, respectively.
Overall, both modules were found to contribute to the performance improvement, especially the significant improvement caused by bias mitigation in the SE block.

The same trend was observed in the case of CAM++.
The original guided speaker embedding (\system{B2}{2}) improved the EERs of the single-speaker baseline (\system{B2}{1}) in all the \svsm{} conditions while degraded in the \svss{} condition.
The proposed bias-mitigated guided speaker embedding (\system{P2}{1}) successfully outperformed \system{B2}{2} in all the cases.
In terms of each module, guided CAM++ is important for performance improvement in all cases, while guided batch normalization contributed to performance improvement in low-overlapping conditions.

To verify the proposed method's effectiveness in mitigating non-target speaker bias, we analyzed EER transitions as non-target-speaker-only durations were scaled like in \autoref{fig:cosine_similarity}.
As shown by each blue line in \autoref{fig:bias_mitigation}, increasing non-target speaker intervals degraded EER with the original guided speaker embedding (\system{B2}{2}).
One might assume that not using intervals where only non-target speakers speak (i.e., $m=0$) would improve performance, but this significantly degrades EER when the overlap ratio is high ($\geq$\SI{75}{\percent}).
This is because information that should not be included in speaker embeddings, easily obtained from intervals containing only non-target speakers, cannot be identified and exploited.
The proposed method eliminated this negative trend, achieving stable EER regardless of non-target-speaker-only duration (red lines in \autoref{fig:bias_mitigation}).

\subsection{Speaker diarization results}

\begin{table}[t]
    \caption{Speaker diarization results in DERs (\%) using speaker embedding extractors based on CAM++.}
    \label{tbl:result_diar}
    \centering
    \sisetup{detect-weight,mode=text}
    \renewrobustcmd{\bfseries}{\fontseries{b}\selectfont}
    \renewrobustcmd{\boldmath}{}
    \newrobustcmd{\B}{\bfseries}
    \scalebox{0.75}{%
    \begin{tabular}{@{}l*{6}{S[table-format=2.2]}@{}}
    \toprule
    &\multicolumn{3}{c}{Local diarization:}&\multicolumn{3}{c@{}}{Local diarization:}\\
    &\multicolumn{3}{c}{pyannote.audio 3.1~\cite{plaquet2023powerset}}&\multicolumn{3}{c@{}}{Oracle}\\\cmidrule(l{\tabcolsep}r{\tabcolsep}){2-4}\cmidrule(l{\tabcolsep}){5-7}
    Dataset & \multicolumn{1}{c}{\system{B2}{1}} & \multicolumn{1}{c}{\system{B2}{2}} & \multicolumn{1}{c}{\system{P2}{1}}&\multicolumn{1}{c}{\system{B2}{1}} & \multicolumn{1}{c}{\system{B2}{2}} & \multicolumn{1}{c@{}}{\system{P2}{1}}\\\midrule
    AISHELL-4 \cite{fu2021aishell} & 12.03 & 11.86 & \B 11.83 & \B 3.05 & 3.09 & 3.39 \\
    AliMeeting \cite{yu2022m2met} & 21.03 & 21.24 & \B 19.66 & 10.46 & 5.52 & \B 4.35\\
    AMI Mix-Headset \cite{carletta2007unleashing} & \B 17.43 & 17.87 & 17.45 & 6.16 & 4.30 & \B 3.92 \\
    AMI Array channel 1 \cite{carletta2007unleashing} & 21.48 & 21.10 & \B 20.20 & 7.66 & 6.24 & \B 5.57 \\
    MSDWild (few) \cite{liu2022msdwild} & 25.08 & 24.41 & \B 24.40 & 10.00 & 7.69 & \B 6.95 \\
    MSDWild (many) \cite{liu2022msdwild} & 48.80 & \B 48.05 & 49.19 &28.78 & 27.15 & \B 24.67 \\
    DIHARD III \cite{ryant2021third} & \B 21.12 & 21.45 & 21.18 & 6.03 & 6.56  & \B 6.02 \\
    VoxConverse \cite{chung2020spot} & 11.94 & 11.52 & \B 11.08 & 3.20 & \B 2.49 & 2.56 \\\midrule
    Macro average &22.36 & 22.19 & \B 21.87 & 9.42 & 7.88 & \B 7.18 \\
    \bottomrule
    \end{tabular}%
    }
\end{table}

Speaker diarization performance was evaluated on eight domains from six datasets: AISHELL-4~\cite{fu2021aishell}, AliMeeting~\cite{yu2022m2met}, AMI Mix-Headset and the first channel of array~\cite{carletta2007unleashing}, MSDWild~\cite{liu2022msdwild} few- and many-talker sets, DIHARD III~\cite{ryant2021third}, and VoxConverse~\cite{chung2020spot}.
We used the pyannote.audio 3.1~\cite{plaquet2023powerset} pipeline; it first performs local diarization with a sliding window of \SI{10}{\second} width and \SI{1}{\second} shift and then aggregates window-wise results by agglomerative hierarchical clustering of speaker embeddings.
We used the pretrained local diarization model, and the following three speaker embedding extractors based on CAM++ were compared: the single-speaker model (\system{B2}{1}), the original guided speaker embedding model (\system{B2}{2}), and the proposed bias-mitigated model (\system{P2}{1}).
Only the single-speaker intervals were used for extraction in \system{B2}{1}, while all the speech intervals were used in \system{B2}{2} and \system{P2}{1}.
To purely compare the performance of the extractors, we also examined cases where the oracle local diarization results are available.
Diarization error rate (DER) without collar tolerance was used for evaluation.

The results are shown in \autoref{tbl:result_diar}.
Regardless of the local diarization method, the naive guided speaker embedding (\system{B2}{2}) outperformed the single-speaker model overall, and the proposed bias-mitigated guided speaker embedding (\system{P2}{1}) brought additional improvements.
If we get better local diarization, the effect of the proposed approach becomes even larger; the gains achieved by the proposed method (\system{P2}{1}) over the baselines (\system{B2}{1} and \system{B2}{2}) with pyannote.audio local diarization (0.49 and 0.32 points) are smaller than those of the oracle local diarization (2.24 and 0.70 points).
One possible reason is that the guide based on speech activities is not very effective when local diarization contains errors.
Future work will address developing a method that is robust to such errors or that utilizes soft guidance instead of binary guidance.

\section{Conclusion}
In this paper, we revealed that the performance degradation of guided speaker embedding in low-overlapping conditions is due to bias introduced by non-target speaker intervals.
We proposed a bias mitigation method as a countermeasure, in which statistics calculated only from the target speaker's intervals were used instead of global statistics.
Speaker verification and diarization results demonstrated the effectiveness of the proposed method.

\clearpage
\bibliographystyle{IEEEtran}
\bibliography{mybib}

\end{document}

%% file: figs/diagram.tex
\begin{tikzpicture}[semithick,auto,
block_high/.style={
    rectangle,
    draw,
    fill=black!20,
    text centered,
    text width=3.5em,
    rounded corners,
    minimum height=2.5em,
    minimum width=3.5em},
block/.style={
    rectangle,
    draw,
    fill=white,
    align=center,
    minimum height=1.5em,
    minimum width=2.5em,
    font=\footnotesize},
block_attention/.style={
    rectangle,
    draw,
    fill=white,
    text centered,
    minimum height=0.5em,
    minimum width=4em,
    font=\footnotesize},
speech_activity/.style={
    rectangle,
    draw,
    fill=white,
    minimum height=0.8em,
    inner sep=0em},
cell/.style={
    rectangle,
    draw,
    minimum size=0.25em,
    inner sep=0pt},
cross/.style={
    path picture={ 
        \draw[black](path picture bounding box.north) -- (path picture bounding box.south) (path picture bounding box.west) -- (path picture bounding box.east);
    }
},
add/.style={
    draw,circle,cross,minimum size=0.8em,inner sep=0
},
hadamard/.style={
    draw,
    shape=circle,
    minimum size=0.8em,
    inner sep=0pt,
    append after command={
        \pgfextra{
            \draw[black] (\tikzlastnode.north west) -- (\tikzlastnode.south east);
            \draw[black] (\tikzlastnode.north east) -- (\tikzlastnode.south west);
        }
    }
},
label/.style={
    draw=none,
    align=center,
    font=\footnotesize,
    inner sep=0,
    outer sep=0
},
]
\tikzset{>=stealth}
\tikzstyle{joint}=[{circle,inner sep=0pt,minimum size=0.3em,fill=black}]

\definecolor{mycolor1}{HTML}{FFD5E5}
\definecolor{mycolor2}{HTML}{FFAACC}
\definecolor{mycolor3}{HTML}{FF8082}
\definecolor{mycolor4}{HTML}{FF5599}
\definecolor{mycolor5}{HTML}{FF2A7F}
\definecolor{mycolor6}{HTML}{FF0066}
\definecolor{mycolor7}{HTML}{D40055}
\definecolor{mycolor8}{HTML}{AA0044}
\definecolor{mycolor9}{HTML}{D7EEF4}
\definecolor{mycolor10}{HTML}{AFDDE9}
\definecolor{mycolor11}{HTML}{87CDDE}
\definecolor{mycolor12}{HTML}{5FBCD3}
\definecolor{mycolor13}{HTML}{37ABC6}
\definecolor{mycolor14}{HTML}{2C89A0}
\definecolor{mycolor15}{HTML}{216776}
\definecolor{mycolor16}{HTML}{164450}

\newcommand{\coloredVector}[2]{%
    \foreach \y in {1,...,#1} {
        \pgfmathtruncatemacro\colorindex{#2[\y-1]}
        \fill[mycolor\colorindex] (0em,\y*0.25em-0.25em) rectangle (0.25em,\y*0.25em);
    }
}

\def\embedfirst{{%
8,3,6,2,3,6,7,3,
}}

\definecolor{spkblue}{HTML}{0080B1}
\definecolor{spkred}{HTML}{E4002B}
\definecolor{spkgreen}{HTML}{06C755}

\node[inner sep=0] (spec) {\includegraphics[width=4em,height=2em]{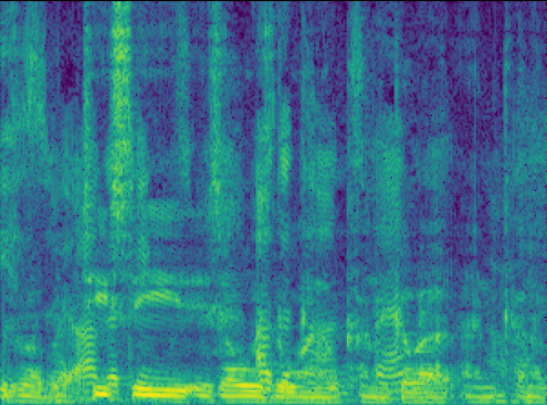}};
\node[label,text width=5.5em,above=0.24em of spec] () {\textbf{Log mel\\filterbank}};

\path let \p1 = ($(spec.south west) + (0.4em,-2.0em)$),
          \p2 = ($(spec.south west) + (2.8em,-2.8em)$) in
    [pattern={Lines[angle=-45,distance=1.5pt]},pattern color=spkred] (\p1) rectangle (\p2);
\node[speech_activity,draw,minimum width=2.4em,anchor=north west,shift={(0.4em,-2.0em)},fill=none] at (spec.south west) (act_target) {};
\path let \p1 = ($(spec.south west) + (0em,-2.8em)$),
          \p2 = ($(spec.south west) + (0.8em,-3.6em)$) in
    [pattern={Lines[angle=45,distance=1.5pt]},pattern color=spkblue] (\p1) rectangle (\p2);
\node[speech_activity,draw,minimum width=0.8em,anchor=north west,shift={(0em,-2.8em)},fill=none] at (spec.south west) (act_nontarget) {};
\path let \p1 = ($(spec.south east) + (0em,-2.8em)$),
          \p2 = ($(spec.south east) + (-2.1em,-3.6em)$) in
    [pattern={Lines[angle=45,distance=1.5pt]},pattern color=spkblue] (\p1) rectangle (\p2);
\node[speech_activity,draw,minimum width=2.1em,anchor=north east,shift={(0em,-2.8em)},fill=none] at (spec.south east) (act_nontarget2) {};
\node[draw=none,align=center,left=0.2em+0.4em of act_target,font=\scriptsize,inner sep=0] (label_target) {Target};
\node[draw=none,align=center,left=0.2em of act_nontarget,font=\scriptsize,inner sep=0] (label_nontarget) {Non-target};
\node[draw=none,fit=(act_target) (act_nontarget) (act_nontarget2) ,inner sep=0.2em] (act) {};
\node[label,text width=8em,above=0.1em of act,shift={(-0.8em,0em)}] () {\textbf{Speech activity}};

\node[add,draw=black,right=0.7em of spec] (concat) {};
\draw[->,thin] (spec) -- (concat);
\draw[->,thick,dash dot] ($(act.east) + (0em,0.3em)$) -| (concat);
\node[draw=none,align=center,text width=4em,rotate=90] at ($(concat.south) + (0.5em,-1.7em)$) () {\footnotesize{Concat.}};
\node[block,right=2.8em of concat,anchor=center,yshift=-0.5em] (encoder) {Encoder\\[-1pt]$f$};
\draw[->,ultra thick,color=spkred] (act.east) -| ($(encoder.south) + (-0.3em,0em)$);
\node[draw=none,align=left,text width=4em,color=spkred] at ($(encoder.south) + (2.1em,-1.5em)$) () {\footnotesize{Bias\\[-3pt]mitigation}};
\draw[->,thin] (concat) -- (concat -| encoder.west);
\node[block,right=7.0em of concat,anchor=center,minimum width=3.4em,fill=none,yshift=-0.5em] (pooling) {Pooling\\[-1pt]$g$};
\draw[->,thin] (concat -| encoder.east) -- (concat -| pooling.west);

\node[draw,dashed,fit=(encoder) (pooling),inner sep=0.2em] (extractor) {};
\node[label,text width=8em,above=0.34em of extractor] () {\textbf{Speaker embedding\\extractor}};

\draw[->,thick,dash dot] ($(act.east) + (0em,-0.3em)$) -| ($(pooling.south) + (0.2em,0em)$);
\node[draw=none,align=center,text width=4em] at ($(pooling.south) + (1.4em,-1.85em)$) () {\footnotesize{Mask}};

\node[rectangle,draw,text width=1*0.25em,
        text height=7*0.25em,inner sep=0,right=14.2em of spec.east,
        path picture={
            \coloredVector{7}{\embedfirst};
            \draw[step=0.25em, black] (path picture bounding box.south west) grid (path picture bounding box.north east);
        }
    ] (e1) {};
    \node[label,text width=7.5em,above=0.2em of e1] () {\textbf{Target speaker's embedding}};
\draw[->,thin] (concat -| pooling.east) -- (e1);

\draw[->,thin] ($(label_nontarget.west) + (0.4em,-1.65em)$) -- ($(label_nontarget.west) + (1.4em,-1.65em)$);
\node[draw=none,align=left,anchor=west,inner sep=0.1em] at ($(label_nontarget.west) + (1.4em,-1.58em)$) (standard) {\footnotesize{: Standard extraction}};
\draw[->,thin] ($(standard.east) + (1em,0em)$) -- ($(standard.east) + (2em,0em)$);
\node[draw=none,align=left,anchor=base west,inner sep=0.1em] at ($(standard.base east) + (2em,0em)$) () {\footnotesize{\&}};
\draw[->,thick, dash dot] ($(standard.east) + (3em,0em)$) -- ($(standard.east) + (4em,0em)$);
\node[draw=none,align=left,text width=40em,anchor=base west,inner sep=0.1em] at ($(standard.base east) + (4em,0em)$) () {\footnotesize{: Guided speaker embedding}};

\draw[->,thin] ($(label_nontarget.west) + (0.4em,-2.65em)$) -- ($(label_nontarget.west) + (1.4em,-2.65em)$);
\node[draw=none,align=left,anchor=west,inner sep=0.1em] at ($(label_nontarget.west) + (1.4em,-2.65em)$) (and1) {\footnotesize{\&}};
\draw[->,thick,dash dot] ($(and1.east) + (0em,0em)$) -- ($(and1.east) + (1em,0em)$);
\node[draw=none,align=left,text width=40em,anchor=west,inner sep=0.1em] at ($(and1.east) + (1em,0em)$) () {\footnotesize{\&}};
\draw[->,ultra thick,spkred] ($(and1.east) + (2em,0em)$) -- ($(and1.east) + (3em,0em)$);
\node[draw=none,align=left,anchor=base west,inner sep=0.1em] at ($(and1.base east) + (3em,0.07em)$) () {\footnotesize{: Bias-mitigated guided speaker embedding (proposed)}};

\draw[very thin] ($(label_nontarget.west) + (0.1em,-0.98em)$) rectangle ($(label_nontarget.west) + (25.3em,-3.28em)$);

\end{tikzpicture}

%% file: figs/basic_se.tex
\begin{tikzpicture}[semithick,auto,
cross/.style={
    path picture={ 
        \draw[black](path picture bounding box.north) -- (path picture bounding box.south) (path picture bounding box.west) -- (path picture bounding box.east);
    }
},
add/.style={
    draw,circle,cross,minimum size=0.8em,inner sep=0
},
hadamard/.style={
    draw,
    shape=circle,
    minimum size=0.6em,
    inner sep=0pt,
    fill=white,
},
speech_activity/.style={
    rectangle,
    draw,
    fill=white,
    minimum height=0.6em,
    inner sep=0em
},
small_text/.style={
    inner sep=0pt,
    font=\small,
},
tiny_text/.style={
    inner sep=0pt,
    font=\tiny,
},
],
\tikzstyle{joint}=[{circle,inner sep=0pt,minimum size=0.3em,fill=black}]
\definecolor{spkblue}{HTML}{0080B1}
\definecolor{spkred}{HTML}{E4002B}
\definecolor{spkgreen}{HTML}{06C755}

\def\opacitybasic{{0.25,0.05,0.5,0.15}};

\newcommand{\grid}[3]{%
    \foreach \x in {1,...,#2} {
        \foreach \y in {1,...,#1} {
            \fill[#3] (\x*0.3em-0.3em,\y*0.3em-0.3em) rectangle (\x*0.3em, \y*0.3em);
        }
    }
}
\newcommand{\maskedgrid}[4]{%
    \foreach \x in {1,...,#2} {
        \foreach \y in {1,...,#1} {
            \pgfmathparse{#4[\y-1]};
            \let\opacityvalue=\pgfmathresult;
            \fill[#3,opacity=\opacityvalue] (\x*0.3em-0.3em,\y*0.3em-0.3em) rectangle (\x*0.3em, \y*0.3em);
        }
    }
}

\node[rectangle,draw=none,text width=7*0.3em,
        text height=4*0.3em,inner sep=0,xshift=0em,
        path picture={
            \grid{4}{7}{spkgreen!50};
        }
    ] () {};
\node[rectangle,draw=black,text width=7*0.3em,
        text height=4*0.3em,inner sep=0,
        path picture={
            \draw[step=0.3em,black] (path picture bounding box.south west) grid (path picture bounding box.north east);
        }
    ] (feat) {};

\node[draw=none,align=center,above left=0.9em and 0.3em of feat.west,font=\tiny,inner sep=0,rotate=90] () {$D$-dim};
\node[draw=none,align=center,below=0.1em of feat,font=\tiny,inner sep=0] () {$T$-frame};
\node[rectangle,draw=black,text width=1*0.3em,
        text height=4*0.3em,inner sep=0,above right=-0.25em and 1.75em of feat,
        path picture={
            \grid{4}{1}{spkgreen!50};
            \draw[step=0.3em,black] (path picture bounding box.south west) grid (path picture bounding box.north east);
        }
    ] (gap) {};
\node[rectangle,draw=black,text width=1*0.3em,
        text height=4*0.3em,inner sep=0,right=2em of gap,
        path picture={
            \maskedgrid{4}{1}{black}{\opacitybasic};
            \draw[step=0.3em,black] (path picture bounding box.south west) grid (path picture bounding box.north east);
        }
    ] (weight) {};
\node[rectangle,draw=none,text width=7*0.3em,
        text height=4*0.3em,inner sep=0,right=6.5em of feat,
        path picture={
            \grid{4}{7}{spkgreen!50};
        }
    ] () {};
\node[rectangle,draw=black,text width=7*0.3em,
        text height=4*0.3em,inner sep=0,right=6.5em of feat,
        path picture={
            \maskedgrid{4}{7}{black}{\opacitybasic};
            \draw[step=0.3em,black] (path picture bounding box.south west) grid (path picture bounding box.north east);
        }
    ] (weighted) {};

\draw[->,name=feat_to_weighted] ($(feat.east) + (0.5mm,0mm)$) -- ($(weighted.west) + (-0.5mm,0mm)$);
\node[joint,right=0.5em of feat.east] (point1) {};
\node[hadamard,draw=black] at ($(point1) + (4.75em,0em)$) (scale) {};
\node[circle,inner sep=0pt,minimum size=0.2em,fill=black] at (scale) {};
\draw[->] (weight) -| (scale);
\draw[->,name=feat_to_gap] (point1) |- (gap.west) node[midway,above=0.15em,tiny_text,align=center] {Eq.\\\autoref{eq:squeeze}};
\draw[->,name=gap_to_weight] (gap.east) -- (weight) node[midway,above=0.15em,tiny_text,align=center] {Eq.\\\autoref{eq:weight}};
\draw[->,name=gap_to_weight] (gap.east) -| (scale) node[midway,above=0.15em,tiny_text,align=center] {Eq.\\\autoref{eq:excitation}};

\node[small_text,above=0.3em of feat,anchor=base] () {$X$};
\node[small_text,above=0.3em of gap,anchor=base] () {$\vect{z}$};
\node[small_text,above=0.3em of weight,anchor=base] () {$\vect{s}$};
\node[small_text,above=0.3em of weighted,anchor=base] () {$X'$};

\path let \p1 = ($(feat.south west) + (0.3em,-0.8em)$),
          \p2 = ($(feat.south west) + (1.5em,-1.4em)$) in
    [pattern={Lines[angle=-45,distance=1.5pt]},pattern color=red] (\p1) rectangle (\p2);
\node[speech_activity,draw,minimum width=1.2em,anchor=north west,shift={(0.3em,-0.8em)},fill=none] at (feat.south west) (act_target) {};
\path let \p1 = ($(feat.south west) + (0em,-1.4em)$),
          \p2 = ($(feat.south west) + (0.6em,-2.0em)$) in
    [pattern={Lines[angle=45,distance=1.5pt]},pattern color=spkblue] (\p1) rectangle (\p2);
\node[speech_activity,draw,minimum width=0.6em,anchor=north west,shift={(0em,-1.4em)},fill=none] at (feat.south west) (act_nontarget) {};
\path let \p1 = ($(feat.south west) + (1.2em,-1.4em)$),
          \p2 = ($(feat.south west) + (2.1em,-2.0em)$) in
    [pattern={Lines[angle=45,distance=1.5pt]},pattern color=spkblue] (\p1) rectangle (\p2);
\node[speech_activity,draw,minimum width=0.9em,anchor=north west,shift={(1.2em,-1.4em)},fill=none] at (feat.south west) (act_nontarget2) {};
\node[draw=none,align=center,right=0.2em+0.6em of act_target,font=\scriptsize,inner sep=0] (label_target) {Target};
\node[draw=none,align=center,right=0.2em of act_nontarget2,font=\scriptsize,inner sep=0,yshift=-0.1em] (label_nontarget) {Non-target};

\draw[gray,thin,densely dotted] ($(feat.south west) + (0.03em,0)$) -- ($(act_nontarget.north west) + (0.03em,0)$);
\draw[gray,thin,densely dotted] ($(feat.south east) - (0.03em,0)$) -- ($(act_nontarget2.north east) - (0.03em,0)$);

\end{tikzpicture}

%% file: figs/target_se.tex
\begin{tikzpicture}[semithick,auto,
cross/.style={
    path picture={ 
        \draw[black](path picture bounding box.north) -- (path picture bounding box.south) (path picture bounding box.west) -- (path picture bounding box.east);
    }
},
add/.style={
    draw,circle,cross,minimum size=0.8em,inner sep=0
},
hadamard/.style={
    draw,
    shape=circle,
    minimum size=0.6em,
    inner sep=0pt,
    fill=white,
},
speech_activity/.style={
    rectangle,
    draw,
    fill=white,
    minimum height=0.6em,
    inner sep=0em
},
small_text/.style={
    inner sep=0pt,
    font=\small,
},
tiny_text/.style={
    inner sep=0pt,
    font=\tiny,
},
],
\tikzstyle{joint}=[{circle,inner sep=0pt,minimum size=0.3em,fill=black}]
\definecolor{spkblue}{HTML}{0080B1}
\definecolor{spkred}{HTML}{E4002B}
\definecolor{spkgreen}{HTML}{06C755}

\def\opacitybasic{{0.25,0.05,0.5,0.15}};

\newcommand{\grid}[3]{%
    \foreach \x in {1,...,#2} {
        \foreach \y in {1,...,#1} {
            \fill[#3] (\x*0.3em-0.3em,\y*0.3em-0.3em) rectangle (\x*0.3em, \y*0.3em);
        }
    }
}
\newcommand{\maskedgrid}[4]{%
    \foreach \x in {1,...,#2} {
        \foreach \y in {1,...,#1} {
            \pgfmathparse{#4[\y-1]};
            \let\opacityvalue=\pgfmathresult;
            \fill[#3,opacity=\opacityvalue] (\x*0.3em-0.3em,\y*0.3em-0.3em) rectangle (\x*0.3em, \y*0.3em);
        }
    }
}

\node[rectangle,draw=none,text width=4*0.3em,
        text height=4*0.3em,inner sep=0,xshift=-0.15em,
        path picture={
            \grid{4}{4}{spkred!35};
        }
    ] () {};
\node[rectangle,draw=black,text width=7*0.3em,
        text height=4*0.3em,inner sep=0,
        path picture={
            \draw[step=0.3em,black] (path picture bounding box.south west) grid (path picture bounding box.north east);
        }
    ] (feat) {};

\node[rectangle,draw=black,text width=1*0.3em,
        text height=4*0.3em,inner sep=0,above right=-0.25em and 1.75em of feat,
        path picture={
            \grid{4}{1}{spkred!35};
            \draw[step=0.3em,black] (path picture bounding box.south west) grid (path picture bounding box.north east);
        }
    ] (gap) {};
\node[rectangle,draw=black,text width=1*0.3em,
        text height=4*0.3em,inner sep=0,right=2em of gap,
        path picture={
            \maskedgrid{4}{1}{black}{\opacitybasic};
            \draw[step=0.3em,black] (path picture bounding box.south west) grid (path picture bounding box.north east);
        }
    ] (weight) {};
\node[rectangle,draw=none,text width=4*0.3em,
        text height=4*0.3em,inner sep=0,right=6.8em of feat,
        path picture={
            \grid{4}{4}{spkred!35};
        }
    ] () {};
\node[rectangle,draw=black,text width=7*0.3em,
        text height=4*0.3em,inner sep=0,right=6.5em of feat,
        path picture={
            \maskedgrid{4}{7}{black}{\opacitybasic};
            \draw[step=0.3em,black] (path picture bounding box.south west) grid (path picture bounding box.north east);
        }
    ] (weighted) {};

\draw[->,name=feat_to_weighted] ($(feat.east) + (0.5mm,0mm)$) -- ($(weighted.west) + (-0.5mm,0mm)$);
\node[joint,right=0.5em of feat.east] (point1) {};
\node[hadamard,draw=black] at ($(point1) + (4.75em,0em)$) (scale) {};
\node[circle,inner sep=0pt,minimum size=0.2em,fill=black] at (scale) {};
\draw[->] (weight) -| (scale);
\draw[->,name=feat_to_gap] (point1) |- (gap.west) node[midway,above=0.15em,tiny_text,align=center] {Eq.\\\autoref{eq:guided_z}};
\draw[->,name=gap_to_weight] (gap.east) -- (weight) node[midway,above=0.15em,tiny_text,align=center] {Eq.\\\autoref{eq:weight}};
\draw[->,name=gap_to_weight] (gap.east) -| (scale) node[midway,above=0.15em,tiny_text,align=center] {Eq.\\\autoref{eq:excitation}};

\node[small_text,above=0.3em of feat,anchor=base] () {$X$};
\node[small_text,above=0.3em of gap,anchor=base] () {$\vect{z}$};
\node[small_text,above=0.3em of weight,anchor=base] () {$\vect{s}$};
\node[small_text,above=0.3em of weighted,anchor=base] () {$X'$};

\path let \p1 = ($(feat.south west) + (0.3em,-0.8em)$),
          \p2 = ($(feat.south west) + (1.5em,-1.4em)$) in
    [pattern={Lines[angle=-45,distance=1.5pt]},pattern color=red] (\p1) rectangle (\p2);
\node[speech_activity,draw,minimum width=1.2em,anchor=north west,shift={(0.3em,-0.8em)},fill=none] at (feat.south west) (act_target) {};
\path let \p1 = ($(feat.south west) + (0em,-1.4em)$),
          \p2 = ($(feat.south west) + (0.6em,-2.0em)$) in
    [pattern={Lines[angle=45,distance=1.5pt]},pattern color=spkblue] (\p1) rectangle (\p2);
\node[speech_activity,draw,minimum width=0.6em,anchor=north west,shift={(0em,-1.4em)},fill=none] at (feat.south west) (act_nontarget) {};
\path let \p1 = ($(feat.south west) + (1.2em,-1.4em)$),
          \p2 = ($(feat.south west) + (2.1em,-2.0em)$) in
    [pattern={Lines[angle=45,distance=1.5pt]},pattern color=spkblue] (\p1) rectangle (\p2);
\node[speech_activity,draw,minimum width=0.9em,anchor=north west,shift={(1.2em,-1.4em)},fill=none] at (feat.south west) (act_nontarget2) {};

\draw[gray,thin,densely dotted] ($(feat.south west) + (0.33em,0)$) -- ($(act_target.north west) + (0.03em,0)$);
\draw[gray,thin,densely dotted] ($(feat.south east) - (0.63em,0)$) -- ($(act_target.north east) - (0.03em,0)$);

\node[draw=none,align=center,above left=0.9em and 0.3em of feat.west,font=\tiny,inner sep=0,rotate=90] () {$D$-dim};
\node[draw=none,align=center,below=0.1em of feat,font=\tiny,inner sep=0] () {$T$-frame};
\node[draw=none,align=center,right=0.2em+0.6em of act_target,font=\scriptsize,inner sep=0] (label_target) {Target};
\node[draw=none,align=center,right=0.2em of act_nontarget2,font=\scriptsize,inner sep=0,yshift=-0.1em] (label_nontarget) {Non-target};

\end{tikzpicture}

%% file: Template.bbl
\begin{thebibliography}{10}
\providecommand{\url}[1]{#1}
\csname url@samestyle\endcsname
\providecommand{\newblock}{\relax}
\providecommand{\bibinfo}[2]{#2}
\providecommand{\BIBentrySTDinterwordspacing}{\spaceskip=0pt\relax}
\providecommand{\BIBentryALTinterwordstretchfactor}{4}
\providecommand{\BIBentryALTinterwordspacing}{\spaceskip=\fontdimen2\font plus
\BIBentryALTinterwordstretchfactor\fontdimen3\font minus \fontdimen4\font\relax}
\providecommand{\BIBforeignlanguage}[2]{{%
\expandafter\ifx\csname l@#1\endcsname\relax
\typeout{** WARNING: IEEEtran.bst: No hyphenation pattern has been}%
\typeout{** loaded for the language `#1'. Using the pattern for}%
\typeout{** the default language instead.}%
\else
\language=\csname l@#1\endcsname
\fi
#2}}
\providecommand{\BIBdecl}{\relax}
\BIBdecl

\bibitem{kinoshita2021integrating}
K.~Kinoshita, M.~Delcroix, and N.~Tawara, ``Integrating end-to-end neural and clustering-based diarization: Getting the best of both worlds,'' in \emph{Proc. ICASSP}, 2021, pp. 7198--7202.

\bibitem{bredin2023pyannote}
H.~Bredin, ``{pyannote.audio 2.1} speaker diarization pipeline: principle, benchmark, and recipe,'' in \emph{Proc. Interspeech}, 2023, pp. 1983--1987.

\bibitem{plaquet2023powerset}
A.~Plaquet and H.~Bredin, ``Powerset multi-class cross entropy loss for neural speaker diarization,'' in \emph{Proc. Interspeech}, 2023, pp. 3222--3226.

\bibitem{lu2021streaming}
L.~Lu, N.~Kanda, J.~Li, and Y.~Gong, ``Streaming end-to-end multi-talker speech recognition,'' \emph{IEEE Signal Processing Letters}, vol.~28, pp. 803--807, 2021.

\bibitem{lu2022endpoint}
L.~Lu, J.~Li, and Y.~Gong, ``Endpoint detection for streaming end-to-end multi-talker {ASR},'' in \emph{Proc. ICASSP}, 2022, pp. 7312--7316.

\bibitem{sklyar2022multi}
I.~Sklyar, A.~Piunova, X.~Zheng, and Y.~Liu, ``Multi-turn {RNN-T} for streaming recognition of multi-party speech,'' in \emph{Proc. ICASSP}, 2022, pp. 8402--8406.

\bibitem{kanda2022streaming}
N.~Kanda, J.~Wu, Y.~Wu, X.~Xiao, Z.~Meng, X.~Wang, Y.~Gaur, Z.~Chen, J.~Li, and T.~Yoshioka, ``Streaming multi-talker {ASR} with token-level serialized output training,'' in \emph{Proc. Interspeech}, 2022, pp. 3774--3778.

\bibitem{cui2024improving}
C.~Cui, I.~Sheikh, M.~Sadeghi, and E.~Vincent, ``Improving speaker assignment in speaker-attributed {ASR} for real meeting applications,'' in \emph{Proc. Odyssey}, 2024, pp. 99--106.

\bibitem{moriya2025alignment}
T.~Moriya, S.~Horiguchi, M.~Delcroix, R.~Masumura, T.~Ashihara, H.~Sato, K.~Matsuura, and M.~Mimura, ``Alignment-free training for transducer-based multi-talker {ASR},'' in \emph{Proc. ICASSP}, 2025.

\bibitem{kanda2019simultaneous}
N.~Kanda, S.~Horiguchi, Y.~Fujita, Y.~Xue, K.~Nagamatsu, and S.~Watanabe, ``Simultaneous speech recognition and speaker diarization for monaural dialogue recordings with target-speaker acoustic models,'' in \emph{Proc. ASRU}, 2019, pp. 31--38.

\bibitem{medennikov2020targetspeaker}
I.~Medennikov, M.~Korenevsky, T.~Prisyach, Y.~Khokhlov, M.~Korenevskaya, I.~Sorokin, T.~Timofeeva, A.~Mitrofanov, A.~Andrusenko, I.~Podluzhny, A.~Laptev, and A.~Romanenko, ``Target-speaker voice activity detection: a novel approach for multi-speaker diarization in a dinner party scenario,'' in \emph{Proc. Interspeech}, 2020, pp. 274--278.

\bibitem{delcroix2021speaker}
M.~Delcroix, K.~Zmolikova, T.~Ochiai, K.~Kinoshita, and T.~Nakatani, ``Speaker activity driven neural speech extraction,'' in \emph{Proc. ICASSP}, 2021, pp. 6099--6103.

\bibitem{horiguchi2025guided}
S.~Horiguchi, T.~Moriya, A.~Ando, T.~Ashihara, H.~Sato, N.~Tawara, and M.~Delcroix, ``Guided speaker embedding,'' in \emph{Proc. ICASSP}, 2025.

\bibitem{desplanques2020ecapatdnn}
B.~Desplanques, J.~Thienpondt, and K.~Demuynck, ``{ECAPA-TDNN}: Emphasized channel attention, propagation and aggregation in {TDNN} based speaker verification,'' in \emph{Proc. Interspeech}, 2020, pp. 3830--3834.

\bibitem{wang2023cam}
H.~Wang, S.~Zheng, Y.~Chen, L.~Cheng, and Q.~Chen, ``{CAM++}: A fast and efficient network for speaker verification using context-aware masking,'' in \emph{Proc. Interspeech}, 2023, pp. 5301--5305.

\bibitem{koluguri2022titanet}
N.~R. Koluguri, T.~Park, and B.~Ginsburg, ``{TitaNet}: Neural model for speaker representation with {1D} depth-wise separable convolutions and global context,'' in \emph{Proc. ICASSP}, 2022, pp. 8102--8106.

\bibitem{zhao2023pcf}
Z.~Zhao, Z.~Li, W.~Wang, and P.~Zhang, ``{PCF}: {ECAPA-TDNN} with progressive channel fusion for speaker verification,'' in \emph{Proc. ICASSP}, 2023.

\bibitem{liu2023ecapa++}
B.~Liu and Y.~Qian, ``{ECAPA++}: Fine-grained deep embedding learning for {TDNN} based speaker verification,'' in \emph{Proc. Interspeech}, 2023, pp. 3132--3136.

\bibitem{thienpondt2021integrating}
J.~Thienpondt, B.~Desplanques, and K.~Demuynck, ``Integrating frequency translational invariance in {TDNN}s and frequency positional information in {2D} {R}es{N}ets to enhance speaker verification,'' in \emph{Proc. Interspeech}, 2021, pp. 2302--2306.

\bibitem{liu2022mfa}
T.~Liu, R.~K. Das, K.~A. Lee, and H.~Li, ``{MFA}: {TDNN} with multi-scale frequency-channel attention for text-independent speaker verification with short utterances,'' in \emph{Proc. ICASSP}, 2022, pp. 7517--7521.

\bibitem{heo2024next}
H.-J. Heo, U.-H. Shin, R.~Lee, Y.~Cheon, and H.-M. Park, ``{NeXt-TDNN}: Modernizing multi-scale temporal convolution backbone for speaker verification,'' in \emph{Proc. ICASSP}, 2024, pp. 11\,186--11\,190.

\bibitem{hu2018squeeze}
J.~Hu, L.~Shen, and G.~Sun, ``Squeeze-and-excitation networks,'' in \emph{Proc. CVPR}, 2018, pp. 7132--7141.

\bibitem{zhou2021resnext}
T.~Zhou, Y.~Zhao, and J.~Wu, ``{ResNeXt} and {Res2Net} structures for speaker verification,'' in \emph{Proc. SLT}, 2021, pp. 301--307.

\bibitem{wang2022efficienttdnn}
R.~Wang, Z.~Wei, H.~Duan, S.~Ji, Y.~Long, and Z.~Hong, ``{EfficientTDNN}: Efficient architecture search for speaker recognition,'' \emph{IEEE/ACM TASLP}, vol.~30, pp. 2267--2279, 2022.

\bibitem{yu2021cam}
Y.-Q. Yu, S.~Zheng, H.~Suo, Y.~Lei, and W.-J. Li, ``{CAM}: Context-aware masking for robust speaker verification,'' in \emph{Proc. ICASSP}, 2021, pp. 6703--6707.

\bibitem{ioffe2015batch}
S.~Ioffe and C.~Szegedy, ``Batch normalization: Accelerating deep network training by reducing internal covariate shift,'' in \emph{Proc. ICML}, 2015, pp. 448--456.

\bibitem{bjorck2018understanding}
N.~Bjorck, C.~P. Gomes, B.~Selman, and K.~Q. Weinberger, ``Understanding batch normalization,'' in \emph{Proc. NeurIPS}, 2018, pp. 7694--7705.

\bibitem{santurkar2018does}
S.~Santurkar, D.~Tsipras, A.~Ilyas, and A.~Madry, ``How does batch normalization help optimization?'' in \emph{Proc. NeurIPS}, 2018, pp. 2488--2498.

\bibitem{nagrani2020voxceleb}
A.~Nagrani, J.~S. Chung, W.~Xie, and A.~Zisserman, ``{VoxCeleb}: Large-scale speaker verification in the wild,'' \emph{Computer Speech \& Language}, vol.~60, p. 101027, 2020.

\bibitem{deng2022arcface}
J.~Deng, J.~Guo, J.~Yang, N.~Xue, I.~Kotsia, and S.~Zafeiriou, ``{ArcFace}: Additive angular margin loss for deep face recognition,'' \emph{IEEE TPAMI}, vol.~44, no.~10, pp. 5962--5979, 2022.

\bibitem{kingma2015adam}
D.~P. Kingma and J.~Ba, ``Adam: A method for stochastic optimization,'' in \emph{Proc. ICLR}, 2015.

\bibitem{snyder2015musan}
D.~Snyder, G.~Chen, and D.~Povey, ``{MUSAN}: A music, speech, and noise corpus,'' arXiv:1510.08484, 2015.

\bibitem{ko2017study}
T.~Ko, V.~Peddinti, D.~Povey, M.~L. Seltzer, and S.~Khudanpur, ``A study on data augmentation of reverberant speech for robust speech recognition,'' in \emph{Proc. ICASSP}, 2017, pp. 5220--5224.

\bibitem{fu2021aishell}
Y.~Fu, L.~Cheng, S.~Lv, Y.~Jv, Y.~Kong, Z.~Chen, Y.~Hu, L.~Xie, J.~Wu, H.~Bu, X.~Xu, J.~Du, and J.~Chen, ``{AISHELL-4}: An open source dataset for speech enhancement, separation, recognition and speaker diarization in conference scenario,'' in \emph{Proc. Interspeech}, 2021, pp. 3665--3669.

\bibitem{yu2022m2met}
F.~Yu, S.~Zhang, Y.~Fu, L.~Xie, S.~Zheng, Z.~Du, W.~Huang, P.~Guo, Z.~Yan, B.~Ma, X.~Xu, and H.~Bu, ``{M2MeT}: The {ICASSP} 2022 multi-channel multi-party meeting transcription challenge,'' in \emph{Proc. ICASSP}, 2022, pp. 6167--6171.

\bibitem{carletta2007unleashing}
J.~Carletta, ``Unleashing the killer corpus: experiences in creating the multi-everything {AMI Meeting Corpus},'' \emph{Language Resources and Evaluation}, vol.~41, no.~2, pp. 181--190, 2007.

\bibitem{liu2022msdwild}
T.~Liu, S.~Fan, X.~Xiang, H.~Song, S.~Lin, J.~Sun, T.~Han, S.~Chen, B.~Yao, S.~Liu, Y.~Wu, Y.~Qian, and K.~Yu, ``{MSDWild}: Multi-modal speaker diarization dataset in the wild,'' in \emph{Proc. Interspeech}, 2022, pp. 1476--1480.

\bibitem{ryant2021third}
N.~Ryant, P.~Singh, V.~Krishnamohan, R.~Varma, K.~Church, C.~Cieri, J.~Du, S.~Ganapathy, and M.~Liberman, ``The third {DIHARD} diarization challenge,'' in \emph{Proc. Interspeech}, 2021, pp. 3570--3574.

\bibitem{chung2020spot}
J.~S. Chung, J.~Huh, A.~Nagrani, T.~Afouras, and A.~Zisserman, ``Spot the conversation: Speaker diarisation in the wild,'' in \emph{Proc. Interspeech}, 2020, pp. 299--303.

\end{thebibliography}
